\documentclass[a4paper,fleqn,usenatbib]{mnras}

\usepackage[T1]{fontenc}
\usepackage{ae,aecompl}

\usepackage{graphicx}	
\usepackage{amsmath}	
\usepackage{amssymb}	

\usepackage{csquotes}

\title[mini-outbursts of GRS 1739$-$278]
{Detection of X-ray spectral state transitions in mini-outbursts of black hole transient GRS 1739$-$278}

\author[Zhen Yan et al.] {
	  Zhen Yan$^{1}$\thanks{E-mail:zyan@shao.ac.cn},
	   Wenfei Yu$^{1}$\thanks{E-mail:wenfei@shao.ac.cn}
	  \\
          $^{1}$Key Laboratory for Research in Galaxies and Cosmology, Shanghai Astronomical Observatory, Chinese Academy of Sciences, \\ 80 Nandan Road, Shanghai 200030, China.; \\
	}
\begin{document}

\date{Accepted .... Received ; in original form }

\pagerange{\pageref{firstpage}--\pageref{lastpage}} \pubyear{2017}

\maketitle

\label{firstpage}

\begin{abstract}
We report the detection of the state transitions and hysteresis effect in the two mini-outbursts of the black hole (BH) transient GRS 1739$-$278 following its 2014 major outburst. The X-ray spectral evolutions in these two mini-outbursts are similar to the major outburst, in spite of their peak luminosities and the outburst durations are one order of magnitude lower. We found the $L_\mathrm{hard-to-soft}$ and the $L_\mathrm{peak,soft}$ of the mini-outbursts also follow the correlation previous found in other X-ray binaries. The $L_\mathrm{hard-to-soft}$ of the mini-outbursts is still higher than that of the persistent BH binary Cyg X-1, which supports that there is a link between the maximum luminosity a source can reach in the hard state and the corresponding non-stationary accretion represented by substantial rate-of-change in the mass accretion rate during flares/outbursts. The detected luminosity range of these two mini-outbursts is roughly in 3.5$\times 10^{-5}$-- 0.015 ($D/7.5$ kpc)$^{2}$($M/8M_{\sun}$) $L_\mathrm{Edd}$. The X-ray spectra of other BH transients at such low luminosities are usually dominated by a power-law component, and an anti-correlation is observed between the photon index and the X-ray luminosity below 1\% $L_\mathrm{Edd}$. So the detection of X-ray spectral state transitions indicates that the accretion flow evolution in these two mini-outbursts of GRS 1739$-$278 are different from other BH systems at such low-luminosity regime.
\end{abstract}


\section{Introduction}
\label{intro}
Most known stellar-mass black holes (BHs) in our Galaxy are harbored in low mass X-ray binary transients (LMXBTs), which are also called BH transients. After a long quiescent period they occasionally undergo an outburst, which is triggered by the disk instability \citep[see reviews in ][]{Lasota2001}. For most bright outbursts, the luminosity can change by several orders of magnitude, and following a certain pattern the X-ray spectrum evolves distinctively between different states \citep[see ][for reviews]{Remillard2006,Belloni2010,Zhang2013}. During the initial rise phase of the outburst, the source is in the hard X-ray state, in which the X-ray spectrum is dominated by a power-law component with a photon index $\Gamma<2.1$. In this state, the X-ray emission is believed to be from a hot accretion flow \citep[see review in ][]{Yuan2014}. During the peak or the initial decay phase of the outburst, on the other hand, the source is in the soft state where the X-ray spectrum is dominated by a multi-temperature blackbody component, emitted by the standard cold accretion disk \citep{Shakura1973}. The source will return to the hard state during the late decay phase of the outburst. In other words, the source experiences a hard-to-soft state transition during the rise phase of the outburst and a soft-to-hard state transition during the decay phase. The standard model for the state transition is that, there exists a critical mass accretion rate for the hot accretion flow (responsible for the hard state), i.e. when the mass accretion rate is above/under that criteria, the source will leave/return the hard state \citep{Esin1997}. However, the hard-to-soft state transition is usually observed at a higher luminosity compared to that of the soft-to-hard state transition, i.e. $L_\mathrm{hard-to-soft} > L_\mathrm{soft-to-hard}$, and it is usually called a hysteresis effect \citep{Miyamoto1995,Maccarone2003}. One powerful tool to track the flux and spectral evolution during the outburst is through the hardness-intensity diagram (HID, \citealt{Belloni2010}),  which plots the X-ray count rate as a function of count rate ratio between two X-ray energy bands. The state transition as well as the hysteresis effect is clearly demonstrated in the HID diagram, where the outburst with state transitions follows a `q'-track \citep{Belloni2010}. Interestingly, it has been found observationally that the $L_\mathrm{hard-to-soft}$ correlates with the peak luminosity of the following soft state $L_\mathrm{peak, soft}$ in individual sources \citep{Yu2004,Yu2007,Yu2007a} as well as in a sample of X-ray binaries \citep{Yu2009,Tang2011}. These results indicate that the well-known hysteresis effect in state transition is primarily because, the variation of the $L_\mathrm{hard-to-soft}$ is tightly connected to the non-stationary property of accretion, which is mainly characterized by the rate-of-change of the mass accretion rate \citep{Yu2009}.

The typical morphology of the X-ray light curve of an outburst of BH transient is a fast-rise-exponential-decay(FRED) form. Although, many outbursts show a secondary maxima in their light curves. \citet{Chen1997} classified three morphological types of the secondary maxima: glitches, bumps, and mini-outbursts. A glitch is a peak superposed on the exponential decay, which is quite common in BH transients \citep{Tanaka1996,Chen1997}. The bump and mini-outburst are much shorter in duration and lower in amplitude than a major outburst, and are usually superposed at the end of the decay phase. The mini-outbursts usually occur when the source is close to the quiescence. These features have been observed in several BH transients, e.g., A0620$-00$, GRO J0422$+$32, XTE J1650$-$500, XTE J1752$-233$ and MAXI J1659$-152$ \citep{Kuulkers1998,Callanan1995,Shrader1997,Tomsick2004,Russell2012, Jonker2012,Homan2013}, and also in several neutron star (NS) LMXBTs, e.g., SAX J1808.4$-$3658 and XTE J1701$-$407 \citep{Patruno2009,Patruno2016,Degenaar2011}. Interestingly, similar mini-outbursts features have also been observed in WZ Sge type dwarf novae \citep[DN, ][]{Kuulkers1996,Kuulkers2000,Kato2004}. The mechanism of mini-outburst is still a puzzle to astrophysicists. Similar mini-outbursts phenomenon in different accreting systems suggest that they probably relate to the accretion process rather than the central compact object. It is very challenging to explain the mini-outburst under the framework of disk instability model \citep{Dubus2001,Lasota2001}. Other models have been invoked, such as the irradiation on the companion \citep{Hameury2000} and the enhanced viscosity \citep{Osaki2001,Meyer2015}. Although these models are designed to interpret the mini-outbursts of DN, it may be applied to X-ray binaries in general.

GRS 1739$-$278 was discovered by the SIGMA gamma-ray telescope on board the Granat satellite \citep{Paul1996,Vargas1997} during its 1996 outburst. Later on, GRS 1739$-278$ was identified as a BH candidate according to its X-ray spectral and timing properties \citep{Borozdin1998,Borozdin2000,Wijnands2001}. In 2014, the source underwent an outburst followed by at least two mini-outbursts. Here we report on the discovery of X-ray spectral state transitions and hysteresis effect in the two mini-outbursts (see \autoref{decay}).

\section{Observations, Data Analysis and Results}
\subsection{$Swift$/XRT observations and data reduction}
This new outburst of GRS 1739$-$278 was firstly detected by the Burst Alert Telescope (BAT) onboard $Swift$ on 2014 March 9 \citep[MJD 56725, ][]{Krimm2014}. The $Swift$ monitoring started on 2014 March 20 (MJD 56736), but was interrupted since MJD 56962 due to Sun constraints. $Swift$ renewed its monitoring of this source about 200 days later, and it is found to remain active. Two mini-outbursts with lower peak luminosities and short durations were detected.

The $Swift$/XRT event data were firstly processed with {\scriptsize XRTPIPELINE} (v 0.13.2) to generate the cleaned event data. We then used {\scriptsize XSELECT} to extract only grade 0 events. Since GRS 1739$-$278 is near the Galactic center with very high column density, higher grade events are omitted in order to avoid possible calibration issue for the heavily absorbed sources \footnote{\url {http://www.swift.ac.uk/analysis/xrt/digest_cal.php\#abs}}. For the observations with high count rate ($>$ 0.6 c/s for the photon counting (PC) mode and $>$ 150 c/s for the windowed timing (WT) mode), we excluded the events in the central region that suffers the pile-up effect \citep{Evans2009}.

\subsection{XRT spectral analysis during the major outburst \label{sub_major}}
\begin{figure*}
\centering
\includegraphics[width=0.8\linewidth]{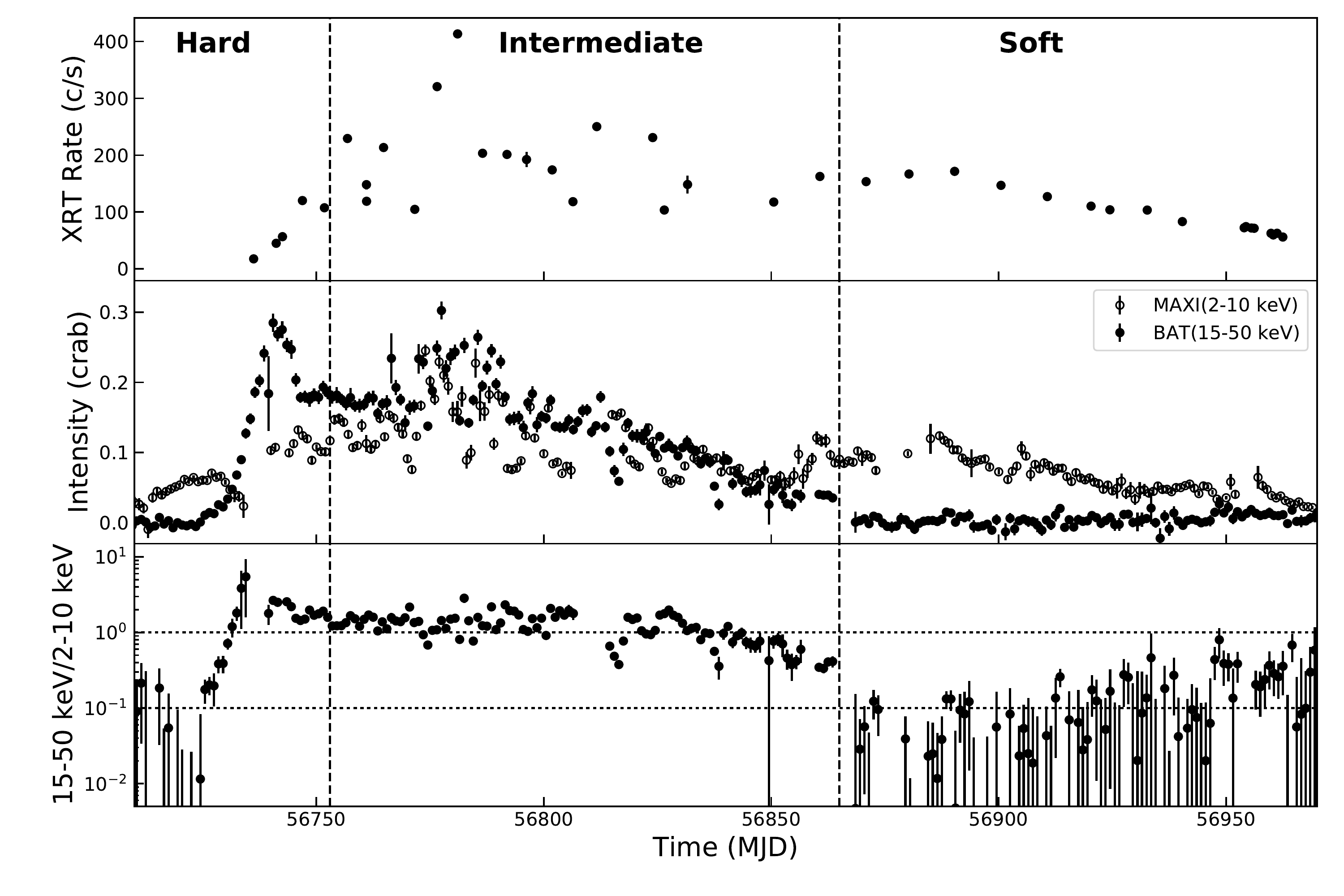}
\caption{The light curves of the major outburst. The intervals of different X-ray spectral states are indicated by the dashed lines. Top panel: The $Swift$/XRT (0.3--10 keV) light curve. Middle panel: The MAXI (2--10 keV) and $Swift$/BAT (15--50 keV) daily light curve. Bottom panel: The hardness ratio between the count rates of BAT and MAXI. The black dotted lines represent the hardness thresholds for the hard and soft states in BH binaries \citep{Yu2009}.}
\label{major}
\end{figure*}

We used {\scriptsize XSPEC} 12.9.0  \citep{Arnaud1996} for the XRT spectral fitting. The first three observations were fitted well with a single power-law component whose photon index $\Gamma\sim$ 1.4 \citep[see also][]{Filippova2014,Miller2015}. Then the photon index increased to $\Gamma\sim$2.0 after MJD 56742. For the data after MJD 56752, an additional disk component was required to fit the XRT spectra. The photon indices were not constrained well in few observations since the power-law component was weak in the energy range of $Swift$/XRT. For simplicity, the photon indices in those observations were fixed to 2.4, a  typical value for the intermediate state or steep power-law state \citep{Remillard2006,Belloni2010}. The XRT spectra after MJD 56860 during the major outburst were all well-fitted by a single disk component, and the X-ray flux and the temperature of the inner disk gradually decreased.

\autoref{major} shows the light curves of the major outburst of GRS 1739$-$278 obtained from $Swift$/XRT \citep[0.3--10 keV;][]{Evans2009}, MAXI \citep[2--10 keV; ][]{Matsuoka2009} and $Swift$/BAT\citep[15-50 keV; ][]{Krimm2013}. The spectral states intervals are defined according to \citet{Remillard2006,Belloni2010}, i.e., the hard state spectrum is characterized by a single power-law with photon index less than 2.1, the soft state spectrum is characterized by a single disk component and the in-between spectra are in the intermediate state (see \autoref{major}). We also plotted the hardness ratio between the count rates of BAT (15--50 keV) and MAXI (2--10 keV) in \autoref{major}. The hardness ratio roughly dropped by a factor of 10 after MJD 56860, mainly because this source entered into a soft state and the hard X-ray intensity by BAT dropped significantly.

\subsection{XRT spectral analysis during the mini-outbursts \label{mini}}
\begin{figure*}
\centering
\includegraphics[width=0.8\linewidth]{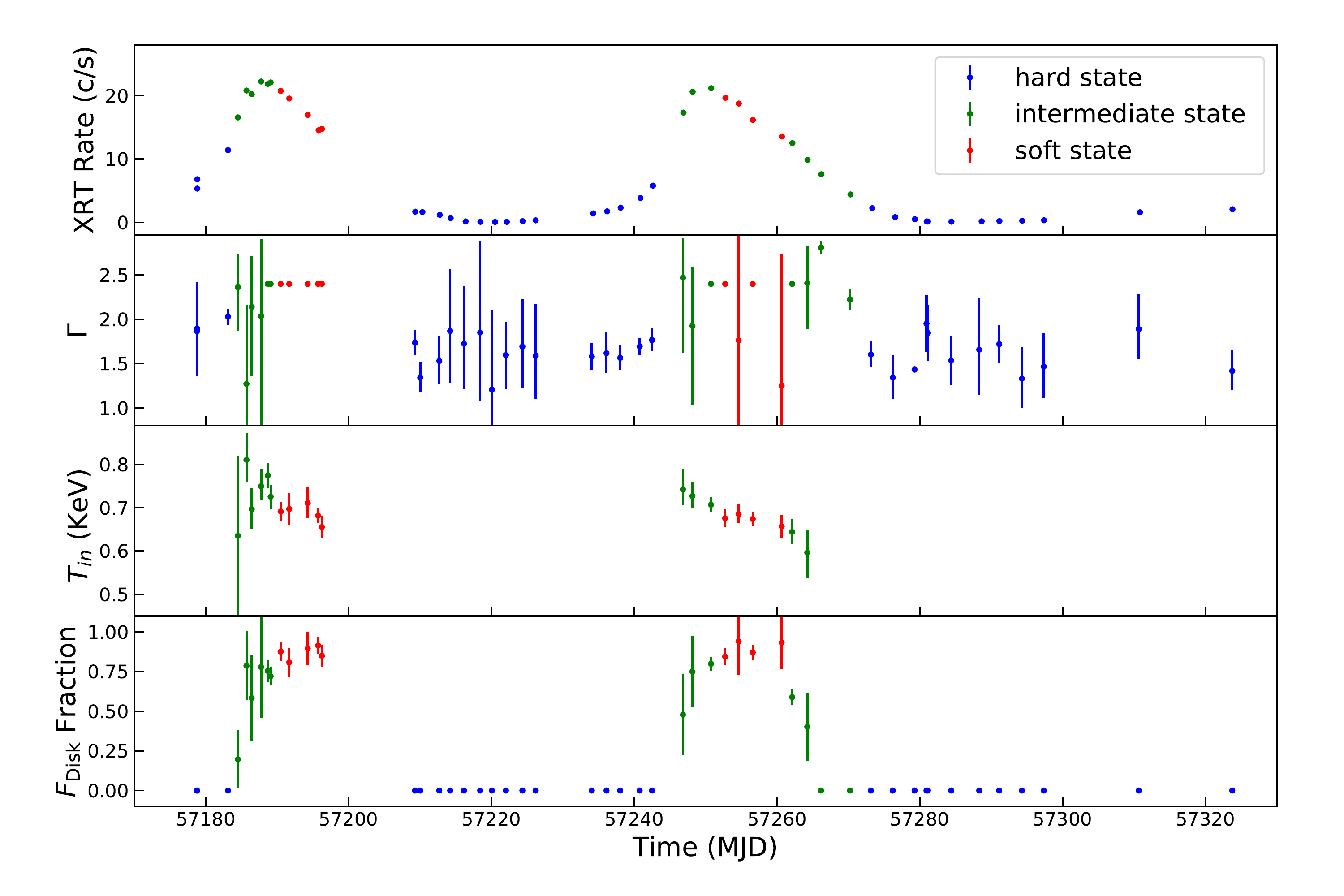}
\caption{The spectral fitting results of the two mini-outbursts. Panels from to to bottom show the  $Swift$/XRT count rate, the power-law photon index $\Gamma$, the inner disk temperature, and the flux fraction from the disk component. As labeled in the figure, different spectral states are indicated by different colors.}
\label{fit}
\end{figure*}

\begin{figure*}
\centering
\includegraphics[width=0.8\linewidth]{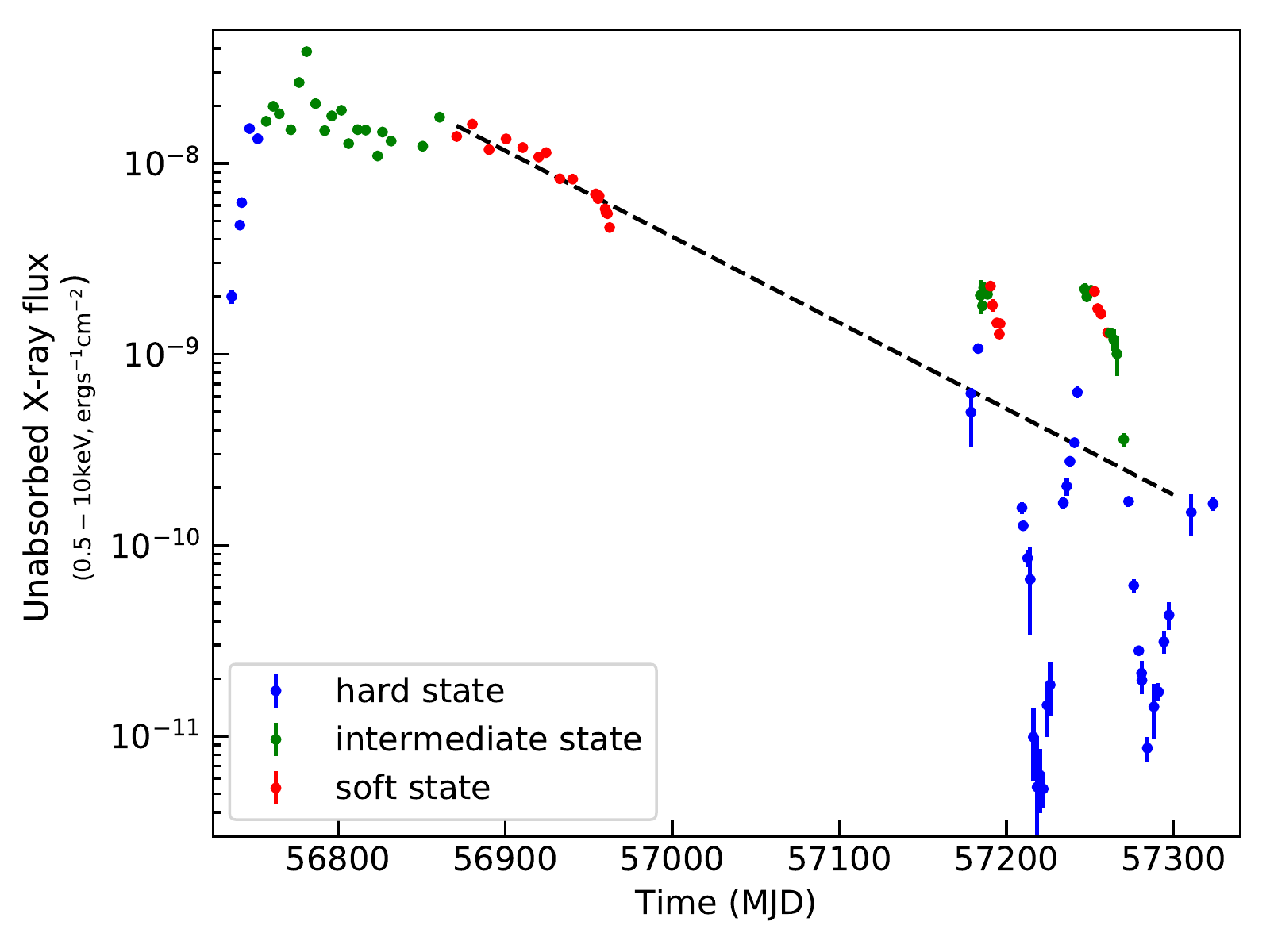}
\caption{The $Swift$/XRT light curve of the major outburst and two mini-outbursts. The decay phase during the soft state follows an exponential form (dashed line). Clearly the mini-outbursts do not superpose on this exponential decay. The spectral states evolution in the two mini-outbursts is similar to the major outburst. Different spectral states are indicated by different colors.}
\label{decay}
\end{figure*}

\begin{figure*}
\centering
\includegraphics[width=0.8\linewidth]{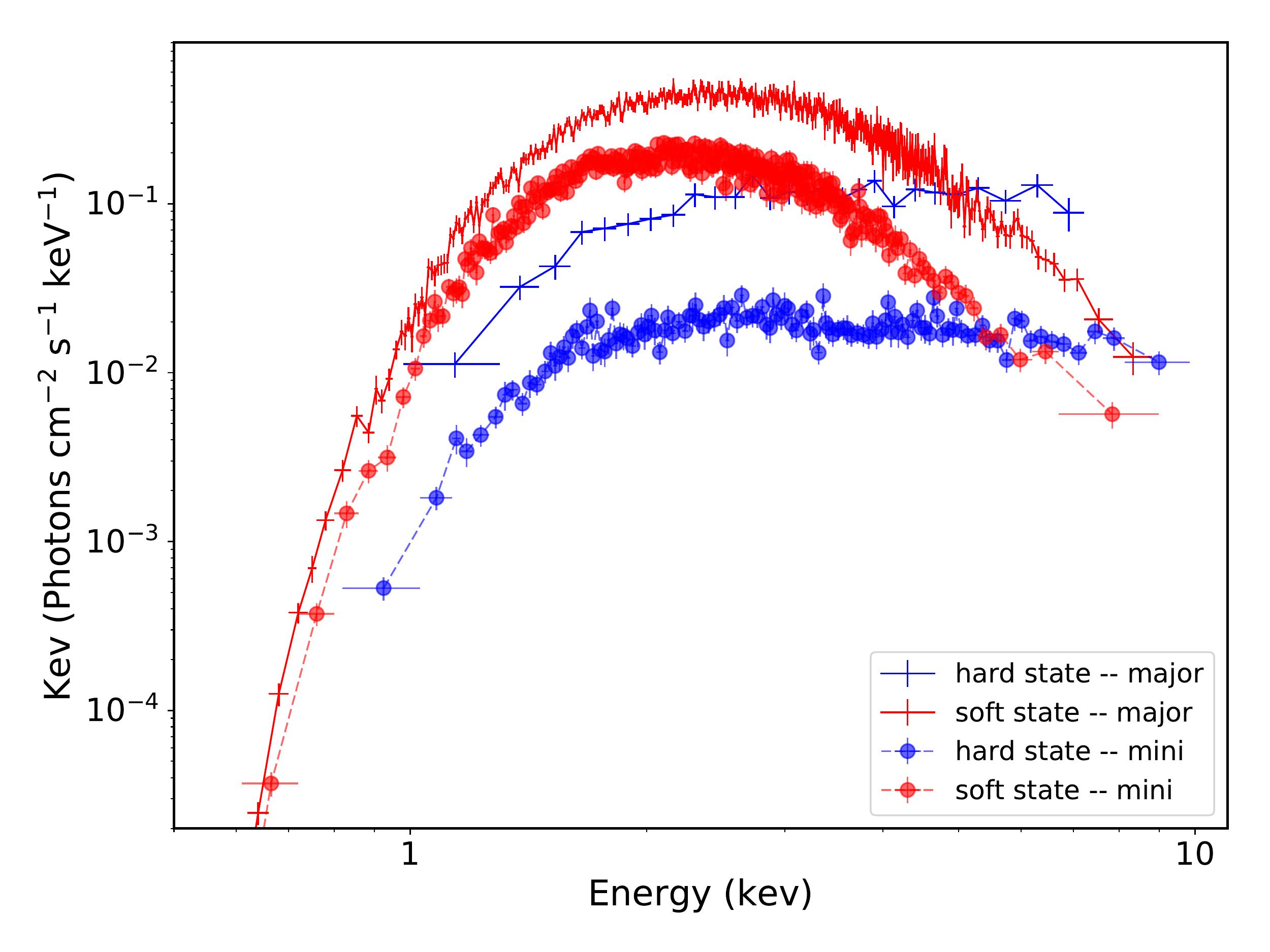}
\caption{A comparison of the $Swift$/XRT unfolded spectra of hard/soft states between the major outburst and mini-outbursts. The spectra of the hard/soft state in the mini-outbursts are quite similar to those in the major outburst.}
\label{spec}
\end{figure*}

\begin{figure*}
\centering
\includegraphics[width=0.8\linewidth]{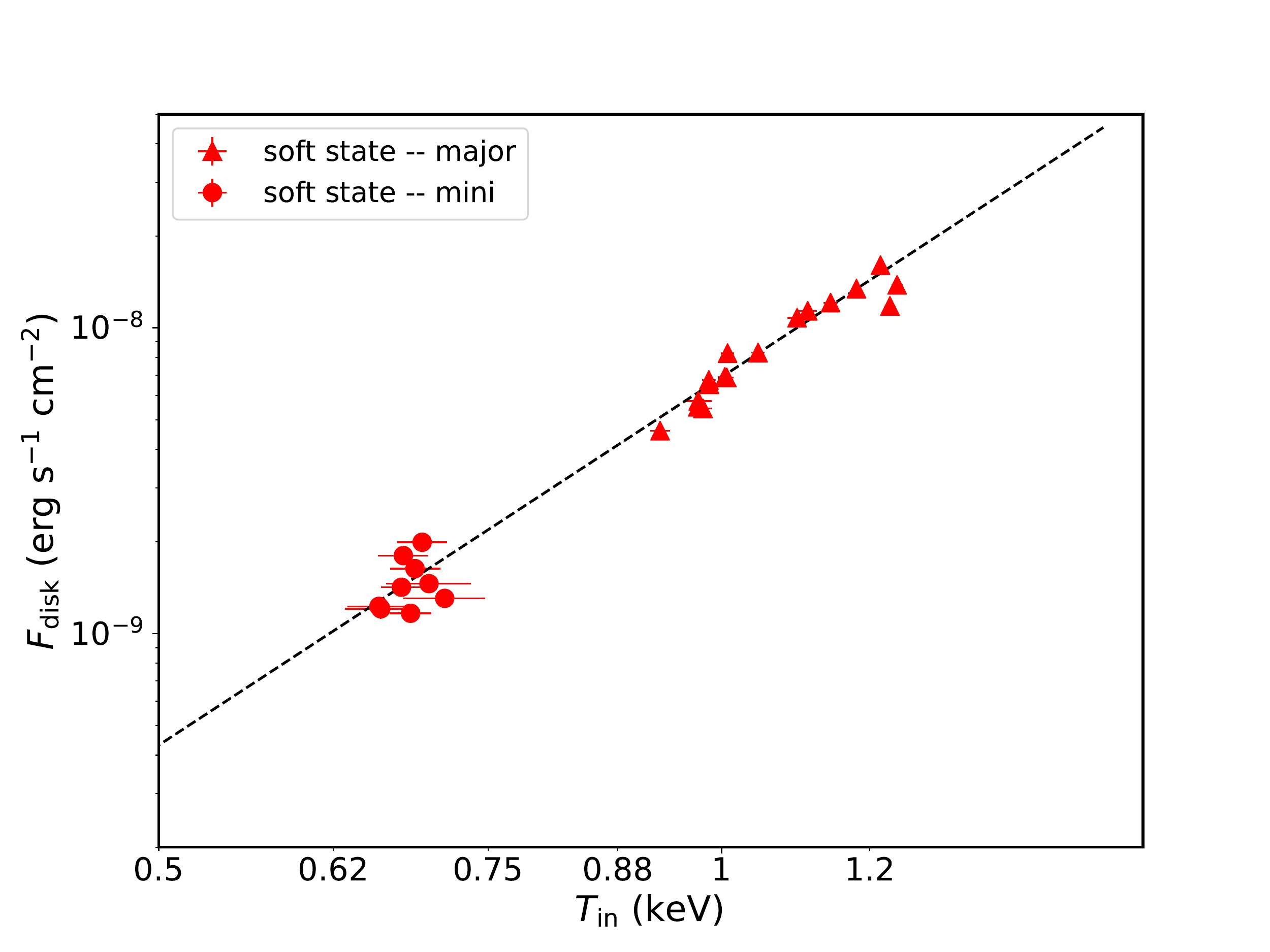}
\caption{The disk flux as a function of the inner disk temperature in the soft states of both major outburst and mini-outbursts of GRS 1739$-$278. The dashed line represents the $F_\mathrm{disk} \propto T_\mathrm{in}^{4}$ relation. The data in the soft states of mini-outbursts roughly follow this relation.}
\label{t_f}
\end{figure*}

The best-fitting parameters of the XRT spectra of GRS 1739$-278$ during the two mini-outbursts are shown in \autoref{fit}. The first $Swift$ observation of the mini-outbursts was performed on MJD 57178, which is about 200 days after the last observation of major outburst. The first few spectra were well fitted by a power-law component. The photon index increased from $\sim$1.8 to $\sim$ 2.1. Then an additional strong disk component was required for the spectral fitting. In some observations, the power-law component was too weak to constrain, and we for simplicity fixed the photon index to 2.4, following  spectral modeling experience of the major outburst. Right after the peak of mini-outbursts, the flux from the disk component could contribute to more than 80\% of the total flux in the 0.5--10 keV (see \autoref{fit}). A single disk component also provided an acceptable fitting for those spectra, but above 6 keV there were still a weak excess, which can be fitted by a power-law component. The $F$-test showed that an additional power-law component improved the spectral fitting at 2--8 $\sigma$ significance level, depending on different observations. During the decay of the mini-outbursts, the X-ray spectra hardened with decreasing flux. The photon index dropped to $\sim$ 1.4--1.5 at the lowest flux level \citep[see also][]{Furst2016}, similar to the initial hard state of the major outburst \citep[see also ][]{Filippova2014,Miller2015}. 

We defined the hard state in the mini-outbursts as the spectra characterized by a single power-law component with  photon index less than 2.1, the soft state as the flux contribution from disk component being larger than 80\% and the in-between spectra are in the intermediate state (see \autoref{fit}). Although their duration and peak luminosity are one oder of magnitude smaller (\autoref{decay}), the two mini-outbursts follow a similar X-ray spectral evolution track to the major outburst and bright outbursts in other BH transients (see \autoref{intro} and the references therein). As shown in the \autoref{decay}, the two mini-outbursts started in the hard state during the rise and made a transition to the soft state after the peak, then returned to the hard state during the decay. We further directly compared the X-ray spectra between major outburst and mini-outbursts (\autoref{spec}). We found that spectra of the hard/soft states in mini-outbursts are almost the same (in spectral shape) as those of the major outburst, except that the inner disk temperature of the soft state in the major outburst is higher than that in the mini-outbursts. Such similarity supports that there are indeed distinct X-ray spectral states and state transitions during the mini-outbursts as well as during the major outburst. 

We then investigated the relationship between the the disk flux $F_\mathrm{disk}$ and the inner disk temperature $T_\mathrm{in}$ during the soft state of mini-outbursts and the major outburst. A simple power-law fitting between $F_\mathrm{disk}$ and $T_\mathrm{in}$ of the soft state in the major outburst indicates that $F_\mathrm{disk}\propto T_\mathrm{in}^{3.84\pm0.35}$, which is consistent with the $F_\mathrm{disk} \propto T_\mathrm{in}^{4}$ expectation of emission from a geometrically thin, optically thick accretion disk with a constant inner radius \citep{Gierlinski2004,Dunn2011}. Besides, the inner disk temperature in the major outburst is typical among other BH transients in their soft states \citep[$\sim$1 keV;][]{Dunn2011,Reynolds2013}. Further, we found that the soft state of the two mini-outbursts roughly follow the extrapolation of the $F_\mathrm{disk} \propto T_\mathrm{in}^{4}$ relation of the major outburst, which indicates that the inner radius of the accretion disk remains constant during the soft states of the major outburst and mini-outbursts. The inner disk temperature of the soft state during the mini-outbursts is lower by a factor of $\sim 2$ than that during the peak of the major outburst, mostly likely due to the decrease in the mass accretion rate of accretion disk \citep{Frank2002}.

\section{Discussion}
\subsection{Distance and BH mass}
\label{s31}
The interstellar extinction along the line of sight of one object can be related to its distance. With an empirical linear correlation between the optical extinction and the column density $N_\mathrm{H}$, we can use the $N_\mathrm{H}$ from X-ray spectral fitting to estimate the distance, provided the column density from the spectra of X-ray binaries is mainly dominated by the interstellar medium \citep{Miller2009}. \citet{Greiner1996} used a similar method to estimate the distance of GRS 1739$-278$ as 6--8.5 kpc. We would like to reexamine the distance measurement by using $N_\mathrm{H}$ from $Swift$/XRT spectral fitting and an updated extinction map. The mean value and the standard deviation of the best-fitting $N_\mathrm{H}$ from all the $Swift$/XRT observations is $2.5\pm 0.5 \times 10^{22}$ cm$^{-2}$, which is in agreement with the individual $XMM-Newton$ and/or $NuSTAR$ observations \citep{Miller2015,Furst2016}. The variation of $N_\mathrm{H}$ in different observations is probably due to the spectral changes especially in the low energy band. Therefore we can constrain the distance of GRS 1739$-278$, if this column density is also dominated by the interstellar medium \citep{Miller2009}. With the up-to-date linear relation between column density and optical extinction $N_\mathrm{H} = (2.87 \pm 0.12) \times 10^{21} A_\mathrm{V}$ cm$^{-2}$  \citep{Foight2016}, we derived the optical extinction $A_\mathrm{V} = 8.7\pm1.8$. Then we converted into near-infrared extinction with the extinction law $A_\mathrm{V}:A_\mathrm{J}:A_\mathrm{H}:A_\mathrm{Ks} = 1:0.188:0.108:0.062$ \citep{Nishiyama2008}. Based on the high resolution 3D near-infrared extinction map of the Galactic bulge \citep{Schultheis2014}, we estimated the distance of GRS 1739$-$278 to be is 5.5--9.5 kpc, consistent with \citet{Greiner1996}.

The BH mass of GRS 1739$-278$ remains unknown. In this work we would like to adopt the mean BH mass of the Galactic BH transients. The mean BH mass of the 12 BH transients with firm mass measurements (cf. \citealt{Corral-Santana2016} for the latest BH transient catalogue) is 8.0$\pm2.0$ $M_{\sun}$ (1 $\sigma$ range). \citet{Ozel2010} also found that the BH mass distribution of transient and persistent BH binaries peaks at 7.8$\pm$1.2 $M_{\sun}$ (see also \citealt{Kreidberg2012}), which roughly agrees with the mean BH mass for the BH transients only.

In the following discussions, we take the distance and the BH mass of GRS 1739$-278$ as 7.5$\pm$2.0 kpc and 8.0$\pm$2.0 $M_{\odot}$, respectively. Correspondingly, the peak luminosity (0.5--10 keV) of the major outburst in 2014 is about 0.25 ($D$/7.5 kpc)$^{2}$ ($M/8M_{\odot}$) $L_\mathrm{Edd}$. The peak luminosity of 1996 outburst is about 0.83 crab in the 2--12 keV \citep[obtained from $RXTE$/ASM and see also][]{Borozdin1998,Borozdin2000}, which is roughly 3.4 times larger than that of 2014 outburst (see \autoref{major}). If we assume there is no spectral change in the two outburst peaks, then the peak luminosity of the 1996 outburst is about 0.85 ($D$/7.5 kpc)$^{2}$ ($M/8M_{\odot}$) $L_\mathrm{Edd}$. The peak luminosities of both outbursts are within the luminosity range observed in the outbursts of BH transients \citep{Reynolds2013,Tetarenko2016}, suggesting that the distance and the BH mass adopted here is reasonable.

\subsection{Implications of the $L_\mathrm{hard-to-soft}$ in the mini-outbursts}

\begin{table*}
	\centering
	\caption{Unabsorbed state transition fluxes and luminosities$^a$ (0.5--10 keV) in major outburst and mini-outbursts}
	\label{l_t}
	\begin{tabular}{lcccc} 
		\hline
		 & $F_\mathrm{hard-to-soft}$ & $F_\mathrm{soft-to-hard}$ & $L_\mathrm{hard-to-soft}$ & $L_\mathrm{soft-to-hard}$ \\
		 &\tiny{($10^{-9}$ erg/s/cm$^{2}$)} & \tiny{($10^{-9}$ erg/s/cm$^{2}$)} &  \tiny{($10^{-2}$ $L_\mathrm{Edd}$)} & \tiny{($10^{-2}$ $L_\mathrm{Edd}$)}  \\
		 
		\hline
		 major outburst  &  13.42$\pm$0.15 to 16.58$\pm$0.31 & $<$4.61 & 8.96$\pm$5.28 to 11.7$\pm$6.52 &  <3.08 \\	
		 first mini-outburst & 1.07$\pm$0.06 to 2.04$\pm$0.41 & 0.16$\pm$0.01 to 1.45$\pm$0.09 & 0.71$\pm$0.42 to 1.27$\pm$0.81 & 0.11$\pm$0.06 to 0.97$\pm$0.57\\
		second mini-outburst &  0.63$\pm$0.05 to 2.20$\pm$0.02 & 0.17$\pm$0.01 to 0.36$\pm$0.03  &0.42$\pm$0.25 to 1.47$\pm$0.87 & 0.11$\pm$0.07 to 0.24$\pm$0.14\\

		\hline
	\multicolumn{5}{l}{ $^a$ The uncertainties on the luminosities have taken into account the uncertainties in distance and BH mass.}  \\	
	\end{tabular}
	
\end{table*}

\begin{figure*}
\centering
\includegraphics[width=0.8\linewidth]{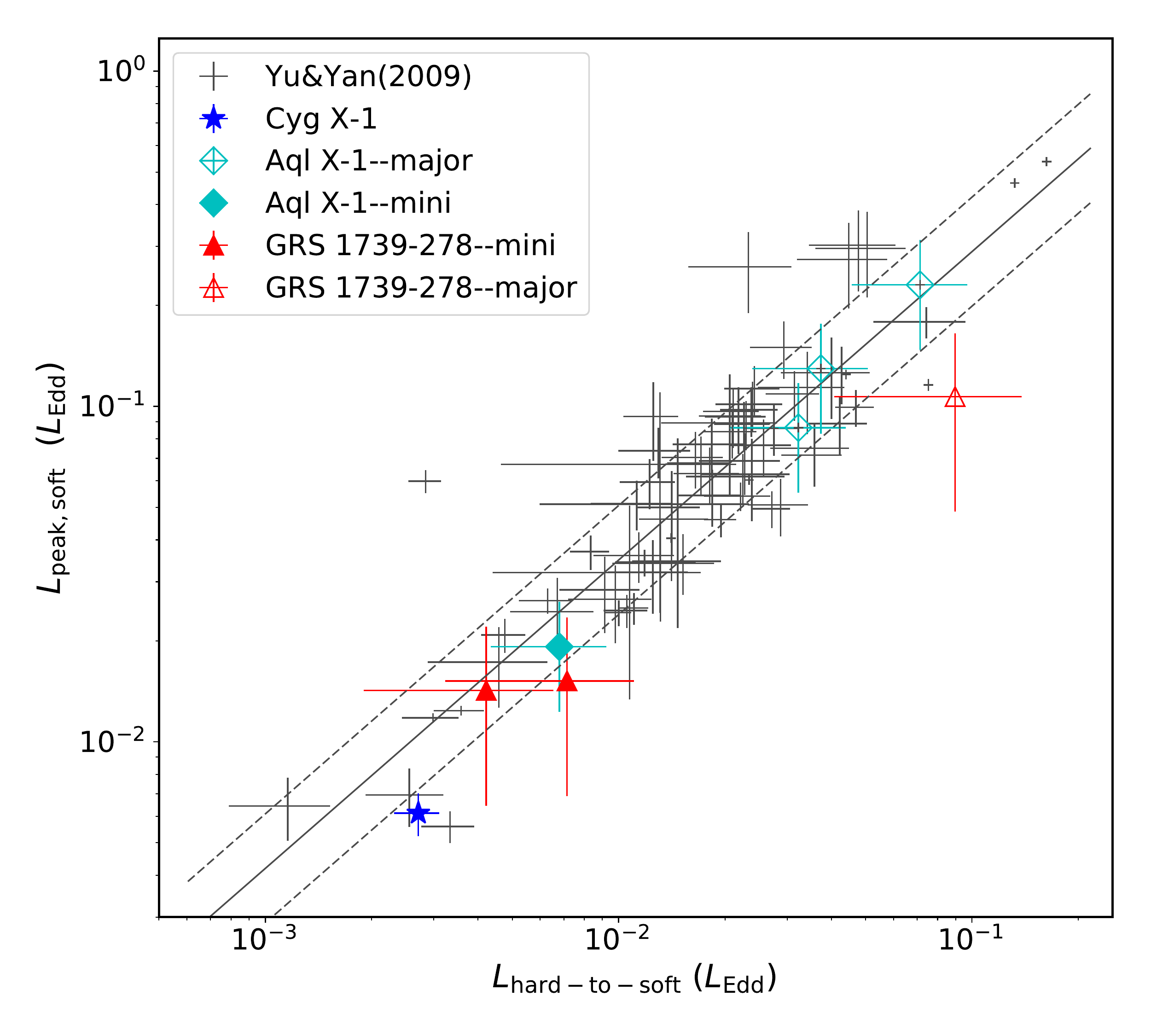}
\caption{The $L_\mathrm{hard-to-soft}$ and $L_\mathrm{peak,soft}$ in the major outburst and mini-outbursts of GRS 1739$-$278 follow the correlation we found in bright XRBs \citep{Yu2009}. The solid line represents the best-fitting function for the sample in \citet{Yu2009}, and the dashed lines show the intrinsic scatters of this correlation.}
\label{f_h2s}
\end{figure*}

Since the X-ray spectral states evolutions in the two mini-outbursts are similar to the the major outburst (see \autoref{mini}), we define the state transition fluxes of the mini-outbursts in the same way to the major outburst. As mentioned in \autoref{intro}, the $F_\mathrm{hard-to-soft}$ represents the flux when the source leaves the hard state during the outburst rise phase and the $F_\mathrm{soft-to-hard}$ represents the flux when the source returns to the hard state during the outburst decay phase. In practice, due to the observational intervals, we did not know the exact time when the source left or returned to the hard state. We thus constrained the $F_\mathrm{hard-to-soft}$ ($F_\mathrm{soft-to-hard}$) to be in the range between that of last (first) hard state during the rise (decay) phase and that of the adjacent subsequent (previous) observation (see \autoref{l_t}). All the unabsorbed state transition fluxes and the corresponding luminosities are listed in \autoref{l_t}. Since $Swift$ observations did not cover the transition to the hard state during the decay phase of the major outburst. The unabsorbed flux of last soft state is about 4.61$\pm 0.08 \times 10^{-9}$ erg s$^{-1}$ cm$^{2}$, which provides an upper limit constraint on $F_\mathrm{soft-to-hard}$ of the major outburst. The $F_\mathrm{soft-to-hard}$ is systematically smaller than $F_\mathrm{hard-to-soft}$, which demonstrates the hysteresis effect \citep[e.g.][]{Miyamoto1995,Maccarone2003} exists in the mini-outbursts as well as in the major outburst.

To our knowledge, this is for the first time to observe the spectral state transitions in mini-outbursts of BH transients (see \autoref{s4}). We note that similar phenomena has been detected in NS X-ray binary Aql X-1 \citep{Yu2007a}, in which a hard-to-soft state transition was detected in a mini-outburst about 200 days after its 2000 outburst. The hard-to-soft state transition luminosity in the mini-outburst is also one order of magnitude lower than that in the major outburst (see \autoref{f_h2s}). But there is no apparent hysteresis effect in the mini-outburst.

An empirical correlation between the transition luminosity $L_\mathrm{hard-to-soft}$ and the peak luminosity of the soft state $L_\mathrm{peak, soft}$ has been found in bright Galactic X-ray binaries \citep{Yu2004,Yu2007,Yu2007a,Yu2009,Tang2011}. It is interesting to check whether the mini-outbursts of GRS 1739$-$278 follow such empirical correlation or not. For this motivation, we used the same method as \citet{Yu2009} to measure the $L_\mathrm{hard-to-soft}$ (i.e., the luminosity of last hard state during the outburst rise phase). The $L_\mathrm{hard-to-soft}$ were measured in 15-50 keV for other sources in \citet{Yu2009}. For the X-ray spectra of GRS 1739$-$278 in the hard state, however, the unabsorbed X-ray flux in 0.5--10 keV is about 2-3 times larger than that in 15--50 keV.  So it seems that there is an offset of $L_\mathrm{hard-to-soft}$ of GRS 1739$-278$ in the \autoref{f_h2s}. The $L_\mathrm{hard-to-soft}$ and $L_\mathrm{peak,soft}$ in mini-outbursts of both GRS 1739$-278$ and Aql X-1 roughly follow the correlation found in bright outbursts (\autoref{f_h2s}), which indicates that the $L_\mathrm{hard-to-soft}$ in the mini-outbursts and the major outburst are likely driven by the same mechanism. The spectral state transition observed in the mini-outbursts suggests that the state transition mechanism can operate at such low luminosity (or a mass accretion rate). Besides, it also indicates that the dynamical range of the mass accretion rate and the timescale for the spectral state transition mechanism to work is at least one order of magnitude. The $L_\mathrm{hard-to-soft}$ of GRS 1739$-278$ in the mini-outbursts are among the lowest transition luminosity detected in BH transients \citep{Gierlinski2006,Yu2009,Dunn2010,Tang2011}. But they are still higher than that of the persistent BH XRB Cyg X-1 (see \autoref{f_h2s}), which supports the schematic picture of the hard state accretion regimes described in Figure 28 of \citet{Yu2009}, and the idea that the maximum luminosity in the hard state $L_\mathrm{hard-to-soft}$ mainly depends on the non-stationary accretion process, which is represented by the substantial rate-of-change in the mass accretion rate during the flare/outburst \citep{Yu2009}.

\subsection{Comparison with mini-outbursts in other BH transients}
\label{s4}
\begin{figure*}
\centering
\includegraphics[width=0.8\linewidth]{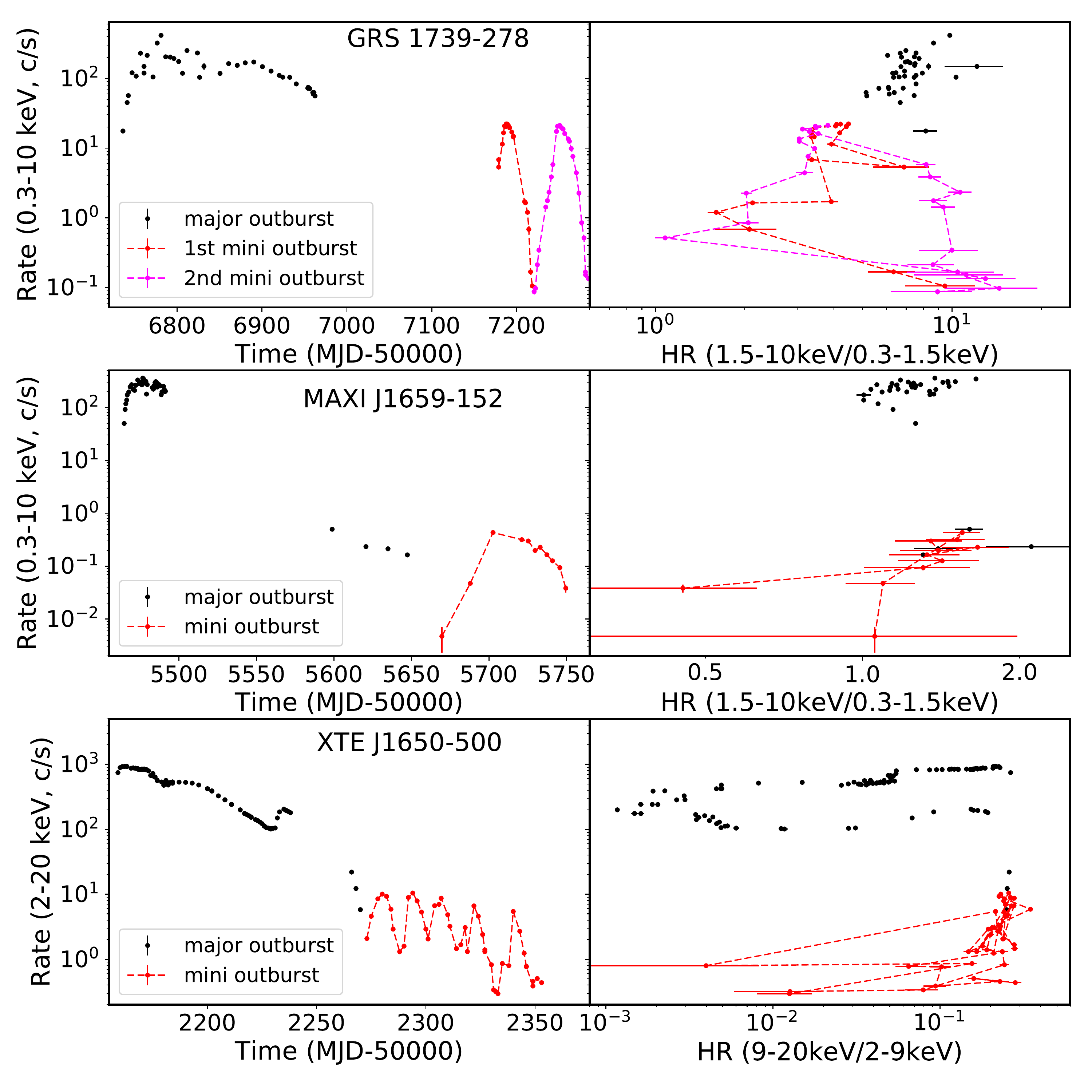}
\caption{Light curves and hardness-intensity diagrams (HIDs) of three BH transients with mini-outbursts. The HIDs of the major outbursts of three sources roughly follow a q-track. The  HIDs of the mini-outbursts of MAXI J1659$-152$ and XTE J16650$-$500 lie on the lower-right of the q-track along with the hard state, and the hardness ratio is likely smaller at the lower intensity. But the HIDs of the mini-outbursts of GRS 1739$-$278 follow the q-track as well as the major outburst.}
\label{qd}
\end{figure*}

There are two mini-outbursts are detected after the 2014 outburst of GRS 1739$-278$. It is unclear whether or not the source went to quiescence after the major outburst due to the observational gap (see \autoref{decay}). The decay phase of the soft state in the major outburst follows an exponential decay profile (see \autoref{decay}). We extrapolated this profile to the end of this outburst, and find that the decay of the first mini-outburst did not resume the previous exponential decay profile. Moreover, considering the long interval between the major outburst and the first mini-outburst ($>$ 200 days), we speculate that two mini-outbursts are not the \enquote{glitches} superposed on the exponential decay \citep{Chen1997}. It is very likely that the first mini-outburst occurred after the end of the major outburst, which are similar to those observed in other BH transients GRO J0422$+$32, XTE J1650$-$500 and MAXI J1659$-152$ \citep{Callanan1995,Shrader1997,Tomsick2004,Homan2005, Jonker2012,Homan2013}.

In order to show the uniquity of the mini-outbursts of GRS 1739$-$278, we compared the X-ray spectral evolutions during the mini-outbursts of these BH transients. First, we excluded the mini-outbursts of GRO J0422$+$32, since they were detected in optical band, while the behavior in X-rays is different to that in the optical band \citep{Shrader1997}. Fortunately, high cadence X-ray observations of the mini-outbursts of XTE J1650$-$500 and MAXI J1659$-152$ had been performed \citep{Tomsick2004,Homan2013}. We plotted in \autoref{qd} the X-ray light curve and HID for these sources. The X-ray count rate and hardness ratio are obtained from $Swift$/XRT online light curve generator \citep{Evans2007,Evans2009} and the standard products of $RXTE$.

The duration and the amplitude of the two mini-outbursts in GRS 1739$-278$ are similar to that observed in MAXI J1659$-152$. But the spectral evolution is quite different. The HIDs of mini-outbursts of MAXI J1659$-152$ and XTE J1650$-500$ mainly lie on the lower-right of the q-track along with the hard state. The hardness ratio is likely smaller at the lower intensity, which agrees with the anti-correlation between the photon index and the luminosity in the low-luminosity regime \citep{Wu2008, Homan2013,Plotkin2013,Cao2014,Yang2015}. The X-ray spectra of MAXI J1659$-152$ and XTE J1650$-500$ during the mini-outbursts are all well-fitted with a single power-law. The photon index range from $\sim$1.5 to $\sim$2.5 for MAXI J1659$-152$ \citep{Homan2013}, and from $\sim$1.7 to $\sim$2.0 for XTE J1650$-500$ \citep{Tomsick2004}, which is consistent with the evolution from hard state towards quiescent state \citep{Plotkin2013}. The HIDs of the mini-outbursts of GRS 1739$-278$ roughly follows a q-track, which are similar to those in the bright outbursts of BH transients \citep[e.g.][]{Dunn2010}. HIDs together with spectral analysis (see \autoref{mini}) support that there are X-ray spectral state transitions and hysteresis effect in the mini-outbursts. For the hard state, the hardness ratio of the mini-outbursts is the same as that of major outbursts (the vertical branch of the q-track). For the soft state, on the other hand, the hardness ratios of the mini-outbursts reach lower range than the major outburst, probably because $T_\mathrm{in}$ is lower in the mini-outburst is lower (see \autoref{t_f}). To conclude, the mini-outbursts in GRS 1739$-278$ look quite similar to bright outbursts in terms of X-ray spectral evolution, while other two BH transients are different.

\subsection{Unique accretion flow evolution in these two mini-outbursts?}
For the two mini-outbursts, the observed lowest 0.5 --10 keV X-ray luminosity is 3.5$\pm$2.2$\times10^{-5}$ $L_\mathrm{Edd}$, and the observed highest luminosity is 0.015$\pm 0.009$ $L_\mathrm{Edd}$, where the BH mass is 8$\pm$2 $M_{\odot}$ and the distance is 7.5$\pm$2.0 kpc. The uncertainties on the luminosities have taken into account the uncertainties in distance and BH mass. Black hole transients in this luminosity range generally stay in the hard state \citep{Dunn2010,Reynolds2013}, with X-ray spectra dominated by a power-law component. Below roughly 0.01 $L_\mathrm{Edd}$, the photon index increases from $\sim 1.4$ to $\sim 2.0$ with decreasing X-ray luminosity \citep[e.g.][]{Tomsick2004,Wu2008,Homan2013,Kalemci2013,Plotkin2013,Cao2014,Yang2015}. This phenomena is also observed in active galactic nuclei \citep[AGN; e.g.][]{Gu2009,Younes2011, Yang2015}. The observed X-ray emission are generally believed to be from the Compton scattering process in hot accretion flow or corona \citep[see reviews in][]{Yuan2014}, although different dynamical models are proposed \citep[e.g.][]{Sobolewska2011,Gardner2013,Qiao2013,Cao2014,Yang2015}.

During the two mini-outbursts of GRS 1739$-$278, the hardest X-ray spectra were detected among the lowest flux levels \citep[$\Gamma\sim1.4$; see also][]{Furst2016}, and the thermal dominated X-ray spectra were detected after the outburst peak. The X-ray spectral evolution during these two mini-outbursts is totally different from other BH transients at such luminosity range, i.e. spectral state transitions are observed. Such state transition behavior usually occurs in the bright outburst with mean peak luminosity $0.2~L_\mathrm{Edd}$ \citep[0.6-- 10 keV;][]{Reynolds2013}, and it is interpreted by the interplay between hot and cold accretion flows  \citep[see reviews in ][]{Done2007,Zhang2013,Yuan2014}. The X-ray emission has different origin in the two states. Interestingly, the $L_\mathrm{hard-to-soft}$ and $L_\mathrm{peak,soft}$ of the mini-outbursts in both BH transient GRS 1739$-$278 and NS transient Aql X-1 follow the same empirical correlation found in bright X-ray binaries (see \autoref{f_h2s}), which indicates that the state transition in mini-outbursts is triggered by the same mechanism as that in bright outbursts. It is still unclear why the state transitions only occur in the mini-outbursts of  GRS 1739$-$278 but not other BH transients. Further investigation, i.e. systematically examining a large sample of mini-outbursts, is in demand to answer this question.

\section*{Acknowledgments}
We are grateful to the anonymous referee for his/her very pertinent comments. We would like to thank Dr. Fu-Guo Xie for helpful discussions. This work was supported in part by the National Key Research and Development Program of China under grant number 2016YFA0400804, the National Natural Science Foundation of China under grant numbers 11403074 and 11333005, and the Strategic Priority Research Program \enquote{The Emergence of Cosmological Structures} under Grant No. XDB09000000. This research has made use of the MAXI data provided by RIKEN, JAXA and the MAXI team, and the Swift data supplied by the UK Swift Science Data Centre at the University of Leicester. This research has made use of data obtained from the High Energy Astrophysics Science Archive Research Center (HEASARC), provided by NASA's Goddard Space Flight Center.

\input{grs1739_clean.bbl}
%



\end{document}